# Kinematical Foundations of Loop Quantum Cosmology

Christian Fleischhack

**Abstract.** First, we review the $C^*$-algebraic foundations of loop quantization, in particular, the construction of quantum configuration spaces and the implementation of symmetries. Then, we apply these results to loop quantum gravity, focusing on the space of generalized connections and on measures thereon. Finally, we study the realm of homogeneous isotropic loop quantum cosmology: once viewed as the loop quantization of classical cosmology, once seen as the symmetric sector of loop quantum gravity. It will turn out that both theories differ, i.e., quantization and symmetry reduction do not commute. Moreover, we will present a uniqueness result for kinematical measures. These last two key results have originally been due to Hanusch; here, we give drastically simplified and direct proofs.

## 1. Introduction

The quantization of gravity is one of the grand unsolved problems in mathematical physics. As can be seen from this edited volume again, there are several different attempts to disclose at least glimpses of the desired theory. Loop quantum gravity [7, 52], the approach that sets the framework for the present contribution, is one of them. It originates in a formulation of gravity, by means of so-called Ashtekar variables [3], as a canonical gauge field theory with constraints that is quantized along the ideas of Dirac. [6, 11] The kinematical part of loop quantum gravity has been well understood; there are even very strong uniqueness results [30, 44] that play the same rôle as the Stone-von Neumann theorem for quantum mechanics. However, despite the marvelous results obtained by Thiemann et al. ([52] and references therein), our knowledge about dynamical consequences of the theory is still rather poor. What to do?

One strategy to investigate complicated physical theories is to study toy models with special symmetries hoping that these models exhibit key features of the full system. In particular, gravity has very much profited from such ideas. Among the first solutions to Einstein's equations, there have been the Schwarzschild solution corresponding to spherically symmetric black holes or the Friedmann solutions that describe homogeneous and isotropic universes. All these have been results within classical gravity. But also in loop quantum gravity, the reduce-symmetry strategy has been successfully implemented. Some 15 years ago, Bojowald, based on his idea with Kastrup [20], invented a theory, now called loop quantum cosmology [4]. He simply quantized cosmological models along the lines of the full theory. It brought about many interesting results like the resolution of big-bang singularities [18] or later big-bounce scenarios [12]. As one had hoped, the dynamics of loop quantum cosmology was much easier accessible, indeed.

The relation between loop quantum gravity and loop quantum cosmology, however, has remained unclear. Before explaining this, let us recall some key ingredients of the loop approach [6, 10, 11, 24]. First, there is some quantum configuration space $\overline{\mathcal{X}}$, which is compact and contains the classical one as a dense subset. Second, $\overline{\mathcal{X}}$ admits some appropriate measure $\mu$ that induces the kinematical $L^2$ Hilbert space. Finally, states are constructed using Gelfand triples. Now, the original idea of Bojowald and Kastrup [20] was as follows: consider the cosmological configuration space as a subset of that of full gravity, and continuously extend this embedding to the corresponding quantum configuration spaces. Based on this extension, one constructs the algebras of basic variables and finally the states by duality.

This connection, however, broke down when Brunnemann and the author [21] discovered that there is *no* continuous extension of the embedding at classical level to one at quantum level. Note that this does not mean that loop quantum gravity or loop quantum cosmology considered as theories of their own are now wrong; it has merely been the bridge between both theories that has been destroyed.

In order to fill that conceptual gap, the kinematics of loop quantum cosmology has had to be carefully revisited. In the present article, we would like to report on the present status of this endeavour [24, 31, 38]. For this, we will first outline the mathematical theory of how to construct quantum configuration spaces in the loop quantization framework, then apply it to loop quantum gravity and finally discuss all the (kinematical) issues above for homogeneous isotropic cosmologies, being in some sense the easiest example. It will surprisingly turn out that the original idea of Bojowald-Kastrup remains valid if one modifies their quantum configuration space appropriately.

## 2. Quantization and Symmetries

The configuration space of loop quantum gravity has originally been obtained using $C^*$-algebras [6]. Later, it has been shown that it can also be identified with a certain projective limit [8, 9]. Although the latter framework is better suited for the construction of measures, the $C^*$-algebraic approach has turned out best suited for our purposes, namely quantization *and* implementation of symmetries.

Its basic idea is as follows: One starts with an arbitrary classical configuration space $\mathcal{X}$. Then, one chooses some $C^*$-algebra $\mathfrak{A}$ consisting of bounded complex-valued functions on $\mathcal{X}$ that are to be quantized in some sense; this is the core part of loop quantization. To implement symmetries, as usual, one assumes to be given a symmetry group $S$ acting[1] on $\mathcal{X}$. This allows to perform

- **Quantization**
  The quantum configuration space $\overline{\mathcal{X}}$ is simply the spectrum of $\mathfrak{A}$.
- **Symmetry Reduction**
  The classical reduced configuration space $\mathcal{X}_{\mathrm{red}}$ comprises the elements in $\mathcal{X}$ that are invariant w.r.t. the symmetry group $S$.

One is now immediately tempted to ask:

- **Quantization ∘ Symmetry Reduction**
  Can we quantize the reduced classical configuration space as the full one? And how are the quantum configuration spaces of the full theory and the reduced theory related? – Well, it will turn out, that the answer crucially depends on the choice of the algebra in the symmetric case.
- **Symmetry Reduction ∘ Quantization**
  Can we implement symmetries even on quantum level? – Yes, provided the underlying algebra is invariant.

This leads us to the most important question: Does it matter whether we implement symmetry *before* or *after* quantization? Or even shorter:

> *Do quantization and symmetry reduction commute?*

Unfortunately, there is no general answer. But at least for some particular cases in loop quantum cosmology, the answer is known as we will have learned by the end of this review.

To start with the general framework, in Subsection 2.1, we review the definition of quantum configuration spaces from [24][2]. Then, in Subsection 2.2 we review, again from [24], a criterion guaranteeing that a mapping between classical configuration spaces can be lifted to their quantum counterparts. The lifting is functorial (Subsection 2.3), which allows to lift even group actions (Subsection 2.4). Finally, in Subsection 2.5, we review the results from [38] on the relations between the quantum symmetric and the symmetric quantum configuration spaces.

### 2.1. Quantum Configuration Spaces

Recall [46] that the spectrum of an abelian $C^*$-algebra $\mathfrak{A}$ consists of all nontrivial multiplicative linear functionals on $\mathfrak{A}$. The Gelfand transform $\widetilde{a}$ of any $a \in \mathfrak{A}$ is given by

$$\widetilde{a}: \begin{array}{ccc} \mathrm{spec}\,\mathfrak{A} & \longrightarrow & \mathbb{C}\,. \\ \chi & \longmapsto & \chi(a) \end{array}$$

---

[1] One may assume w.l.o.g. that the action is effective although we will never use this.
[2] Parts of Proposition 1 originate in [47].

The topology on $\operatorname{spec}\mathfrak{A}$ is the initial topology induced by all the Gelfand transforms. The celebrated Gelfand-Naimark theorem states that $a \longmapsto \widetilde{a}$ is an isometric $*$-isomorphism between $\mathfrak{A}$ and $C_0(\operatorname{spec}\mathfrak{A})$. If now $\mathfrak{A}$ is an algebra of functions on some configuration space $\mathcal{X}$, then the evaluation mapping provides us with a natural embedding of $\mathcal{X}$ into $\operatorname{spec}\mathfrak{A}$.

**Proposition 1.** *Let $\mathcal{X}$ be a set and $\mathfrak{A} \subseteq \ell^\infty(\mathcal{X})$ a $C^*$-subalgebra with empty kernel[3]. Then the **natural mapping** $\iota : \mathcal{X} \longrightarrow \operatorname{spec}\mathfrak{A}$ defined by the evaluation mapping*

$$\iota(x): \quad \mathfrak{A} \quad \longrightarrow \quad \mathbb{C}$$
$$\phantom{\iota(x):\quad} a \quad \longmapsto \quad a(x)$$

*is well defined and has the following properties:*

1. *$\widetilde{a} \circ \iota = a$ for all $a \in \mathfrak{A}$.*
2. *$\iota(\mathcal{X})$ is dense in $\operatorname{spec}\mathfrak{A}$.*
3. *$\iota$ and $\mathfrak{A}$ separate the same points.*
4. *$\iota$ is injective $\iff \mathfrak{A}$ separates the points in $\mathcal{X}$.*
5. *$\iota$ is continuous $\iff \mathfrak{A}$ consists of continuous functions on $\mathcal{X}$ only.*

*Here, for the final assertion, we assumed $\mathcal{X}$ to be given some topology.*

*Proof.* Well-definedness follows from the empty-kernel assumption.

1. Observe $[\widetilde{a} \circ \iota](x) \equiv \widetilde{a}(\iota(x)) = [\iota(x)](a) = a(x)$.
2. Let $\chi \in \operatorname{spec}\mathfrak{A}$ be outside the closure of $\iota(\mathcal{X})$. Then, by regularity of locally compact Hausdorff spaces, there is some $\phi = \widetilde{a} \in C_0(\operatorname{spec}\mathfrak{A})$ with $\phi(\chi) \neq 0$, but $\phi \equiv 0$ on $\iota(\mathcal{X})$. This implies $a = \widetilde{a} \circ \iota = \phi \circ \iota = 0$ and $\phi \equiv 0$.
3. Observe $\iota(x) = \iota(x') \iff a(x) \equiv \iota(x)(a) = \iota(x')(a) \equiv a(x')$ for all $a \in \mathfrak{A}$.
4. This follows immediately from the previous item.
5. By definition of the Gelfand-Naimark topology, $\iota$ is continuous iff $a \equiv \widetilde{a} \circ \iota : \mathcal{X} \longrightarrow \mathbb{C}$ is continuous for all $a$. □

**Definition 2.** *The **quantum configuration space** $\overline{\mathcal{X}}$ is the spectrum of $\mathfrak{A}$.*

We will refer to $\mathfrak{A}$ as the corresponding **quantizing algebra**. This algebra should be rich enough to separate the points of $\mathcal{X}$ and to contain sufficiently many physically "interesting" functions. Note, however, that the definition of $\overline{\mathcal{X}}$ crucially depends on the choice of $\mathfrak{A}$. This will become relevant in loop quantum cosmology.

### 2.2. Lifting Criteria

In the next step, we are going to compare different quantum configuration spaces. More concretely, we assume to be given two classical configuration spaces $\mathcal{X}_1$ and $\mathcal{X}_2$ that are related by some mapping $\sigma : \mathcal{X}_1 \longrightarrow \mathcal{X}_2$. There are two prime examples we will be interested in: First, $\mathcal{X}_1$ is a subset of $\mathcal{X}_2$, given by certain invariant elements in $\mathcal{X}_2$, with $\sigma$ being the canonical embedding; second, both spaces coincide

---

[3]The kernel of $\mathfrak{A}$ is defined by $\bigcap_{a \in \mathfrak{A}} a^{-1}(0)$. In particular, each unital $\mathfrak{A}$ has empty kernel. Throughout the whole article, any $\mathfrak{A}$ will have empty kernel – or we will assume that.

and $\sigma$ is some isomorphism, e.g., obtained from some group action. How can we relate the corresponding quantum configuration spaces? A very natural notion is obviously that of a lift[4]:

**Definition 3.** $\overline{\sigma}$ *is a* **lift** *of* $\sigma$ *(w.r.t. some algebras* $\mathfrak{A}_1$ *and* $\mathfrak{A}_2$*) iff it fills the diagram*

$$\begin{array}{ccc} \mathcal{X}_1 & \xrightarrow{\sigma} & \mathcal{X}_2 \\ \downarrow{\iota_1} & & \downarrow{\iota_2} \\ \overline{\mathcal{X}}_1 & \xrightarrow{\overline{\sigma}} & \overline{\mathcal{X}}_2 \end{array}$$

Is there always a lift $\overline{\sigma} : \overline{\mathcal{X}}_1 \longrightarrow \overline{\mathcal{X}}_2$ filling this diagram? Under which circumstances is $\overline{\sigma}$ unique? When is it continuous? When injective?

To get an idea how to answer these questions, assume that such a continuous lift $\overline{\sigma}$ exists for unital $\mathfrak{A}_1$. Then, for any $a_2 \in \mathfrak{A}_2$, the function $\widetilde{a}_2 \circ \overline{\sigma} : \overline{\mathcal{X}}_1 \longrightarrow \mathbb{C}$ is continuous, hence equals $\widetilde{a}_1$ for some $a_1 \in \mathfrak{A}_1$ by Gelfand-Naimark (and unitality of $\mathfrak{A}_1$). Moreover,

$$\begin{array}{ccccc} \mathcal{X}_1 & \xrightarrow{\sigma} & \mathcal{X}_2 & \xrightarrow{a_2} & \mathbb{C} \\ \downarrow{\iota_1} & & \downarrow{\iota_2} & \nearrow{\widetilde{a}_2} & \\ \overline{\mathcal{X}}_1 & \xrightarrow{\overline{\sigma}} & \overline{\mathcal{X}}_2 & & \end{array}$$

commutes, and we have

$$\sigma^* a_2 \equiv a_2 \circ \sigma = \widetilde{a}_2 \circ \overline{\sigma} \circ \iota_1 = \widetilde{a}_1 \circ \iota_1 = a_1 \in \mathfrak{A}_1.$$

This motivates to define the **restriction algebra**[5] of $\mathfrak{A}_2$ w.r.t. $\sigma$ by

$$\sigma^* \mathfrak{A}_2 := \{\sigma^* a_2 \mid a_2 \in \mathfrak{A}_2\} \subseteq \ell^\infty(\mathcal{X}_1).$$

Thus we have shown that $\sigma^* \mathfrak{A}_2 \subseteq \mathfrak{A}_1$ is necessary for the existence of $\overline{\sigma}$. But, could this condition even be sufficient as well?

So, let us assume $\sigma^* \mathfrak{A}_2 \subseteq \mathfrak{A}_1$. Then, by Proposition $1_3$, we have

$$\iota_1(x_1) = \iota_1(x_1') \iff a_1(x_1) = a_1(x_1') \quad \forall a_1$$
$$\implies a_2(\sigma(x_1)) = a_2(\sigma(x_1')) \quad \forall a_2 \iff \iota_2(\sigma(x_1)) = \iota_2(\sigma(x_1'))$$

whence $\widehat{\sigma} := \iota_2 \circ \sigma \circ \iota_1^{-1} : \iota_1(\mathcal{X}_1) \longrightarrow \overline{\mathcal{X}}_2$ is well defined. Moreover, we have

$$(\sigma^* a_2)^\sim \circ \iota_1 = \sigma^* a_2 = \widetilde{a}_2 \circ \iota_2 \circ \sigma = \widetilde{a}_2 \circ \widehat{\sigma} \circ \iota_1$$

for any $a_2 \in \mathfrak{A}_2$. In particular, $(\sigma^* a_2)^\sim$ is a continuous continuation of $\widetilde{a}_2 \circ \widehat{\sigma}$ on $\overline{\mathcal{X}}_1$. As $\iota_1(\mathcal{X}_1)$ is dense in the compact space $\overline{\mathcal{X}}_1$, one shows quickly that $\widehat{\sigma}$ can be extended to a continuous mapping $\overline{\sigma}$ on $\overline{\mathcal{X}}_1$. In particular, it fulfills

$$(\sigma^* a_2)^\sim = \widetilde{a}_2 \circ \overline{\sigma} \qquad (1)$$

---
[4]If the natural mappings $\iota_i$ are injective, this just means that $\overline{\sigma}$ extends $\sigma$.
[5]To explain the term "restriction algebra", assume that $\sigma$ is injective, whence $\mathcal{X}_1$ can be considered as a subset of $\mathcal{X}_2$. Then $\sigma^* \mathfrak{A}_2$ consists just of the restrictions of the functions in $\mathfrak{A}_2 \subseteq \ell^\infty(\mathcal{X}_2)$ to the domain $\mathcal{X}_1$. In order to avoid conflicts with the different notion of pull-back $C^*$-algebras, we will use the notion "restriction algebra" also in the case where $\sigma$ is not injective.

for all $a_2 \in \mathfrak{A}_2$. Altogether, we have derived

**Proposition 4.** *For unital $\mathfrak{A}_1$, there is a continuous lift of $\sigma$ iff $\sigma^*\mathfrak{A}_2 \subseteq \mathfrak{A}_1$.*

Uniqueness of $\overline{\sigma}$ is obvious as $\iota_1(\mathcal{X}_1)$ is always dense. So, given existence, it remains to find a criterion for injectivity of $\overline{\sigma}$. For this, observe first that we obtain from (1) and Proposition $1_1$

$$\overline{x}_1(\sigma^*a_2) = (\sigma^*a_2)^\sim(\overline{x}_1) = [\widetilde{a}_2 \circ \overline{\sigma}](\overline{x}_1) = [\overline{\sigma}(\overline{x}_1)](a_2)$$

for any $\overline{x}_1 \in \overline{\mathcal{X}}_1$ and $a_2 \in \mathfrak{A}_2$, hence

**Lemma 5.** *If $\mathfrak{A}_1$ is unital and $\overline{\sigma}$ exists, then for all $\overline{x}_1 \in \overline{\mathcal{X}}_1$*

$$\overline{\sigma}(\overline{x}_1) = \overline{x}_1 \circ \sigma^*.$$

This now implies

$$\overline{\sigma}(\overline{x}_1) = \overline{\sigma}(\overline{x}_1') \iff \overline{x}_1(\sigma^*a_2) = \overline{x}_1'(\sigma^*a_2) \quad \forall a_2 \in \mathfrak{A}_2.$$

Consequently, $\overline{\sigma}$ is injective iff $\overline{\mathcal{X}}_1 = \operatorname{spec} \mathfrak{A}_1$ separates to points in $\sigma^*\mathfrak{A}_2 \subseteq \mathfrak{A}_1$, which is equivalent to the denseness of $\sigma^*\mathfrak{A}_2$ in $\mathfrak{A}_1$ by the Stone-Weierstraß and the Gelfand-Naimark theorems, provided $\sigma^*\mathfrak{A}_2$ is unital. Altogether, we have

**Theorem 6.** *Let $\mathfrak{A}_1$ and $\mathfrak{A}_2$ be unital. Then*

$$\sigma : \mathcal{X}_1 \longrightarrow \mathcal{X}_2 \text{ has a continuous injective lift} \iff \sigma^*\mathfrak{A}_2 \text{ is dense in } \mathfrak{A}_1$$

*Moreover, the lift is unique if existing.*

Note that we did *not* require $\sigma$ to fulfill any continuity condition. Indeed, we get the continuity of $\overline{\sigma}$ for free from the $C^*$-algebra construction.

For practical purposes, it is often nicer to use the criterion

**Corollary 7.** *Let $\mathfrak{A}_1$ and $\mathfrak{A}_2$ be unital, and let $\mathfrak{B}_2$ be a dense $*$-algebra in $\mathfrak{A}_2$. Then*

$$\sigma : \mathcal{X}_1 \longrightarrow \mathcal{X}_2 \text{ has a continuous injective lift} \iff \sigma^*\mathfrak{B}_2 \text{ is dense in } \mathfrak{A}_1$$

*Proof.* Observe that $\overline{\sigma^*\mathfrak{B}_2}$ always equals $\overline{\sigma^*\mathfrak{A}_2}$, since

$$\sigma^*\mathfrak{B}_2 \subseteq \sigma^*\mathfrak{A}_2 \equiv \sigma^*\overline{\mathfrak{B}_2} \subseteq \overline{\sigma^*\mathfrak{B}_2}.$$

Now the statement follows from Theorem 6 immediately. □

These criteria directly suggest what to do if one wants to embed the quantum configuration space for $\mathcal{X}_1$ into that of $\mathcal{X}_2$, given some $\sigma$. Simply *define* the algebra $\mathfrak{A}_1$ to be the completion of $\sigma^*\mathfrak{A}_2$. We will apply this strategy to the case of loop quantum cosmology in Section 4, where $\mathcal{X}_2$ is the configuration space of full gravity and $\mathcal{X}_1$ that of symmetric configurations describing a certain cosmology. But, before that, we shall discuss how to lift group actions from any classical configuration space to its quantum counterpart.

### 2.3. Functoriality

Having seen how to lift configuration spaces from the classical to the quantum level, but also how to lift mappings between them, it is very natural to ask whether these constructions are in some sense functorial. Indeed, they are. Consider the diagram:

$$\begin{array}{ccccc}
\mathcal{X}_1 & \xrightarrow{\sigma} & \mathcal{X}_2 & \xrightarrow{\tau} & \mathcal{X}_3 \\
\iota_1 \downarrow & & \iota_2 \downarrow & & \iota_3 \downarrow \\
\overline{\mathcal{X}}_1 & \xrightarrow{\overline{\sigma}} & \overline{\mathcal{X}}_2 & \xrightarrow{\overline{\tau}} & \overline{\mathcal{X}}_3
\end{array}$$

Here, our lifting criterion tells us that $\overline{\sigma}$ exists iff $\sigma^*\mathfrak{A}_2 \subseteq \mathfrak{A}_1$ and that $\overline{\tau}$ exists iff $\tau^*\mathfrak{A}_3 \subseteq \mathfrak{A}_2$. Together, we see that the liftability of $\sigma$ and $\tau$ implies that

$$(\tau \circ \sigma)^*\mathfrak{A}_3 \;\equiv\; \sigma^*(\tau^*\mathfrak{A}_3) \;\subseteq\; \sigma^*\mathfrak{A}_2 \;\subseteq\; \mathfrak{A}_1$$

i.e., $\tau \circ \sigma$ is liftable as well. From the uniqueness part, we get immediately that

$$\overline{\tau \circ \sigma} \;=\; \overline{\tau} \circ \overline{\sigma}.$$

Finally, even the embedding criterion is functorial. Indeed, assume that $\sigma$ and $\tau$ lift to embeddings, i.e., $\sigma^*\mathfrak{A}_2$ is dense in $\mathfrak{A}_1$ as well as $\tau^*\mathfrak{A}_3$ is dense in $\mathfrak{A}_2$. Now,

$$\sigma^*\mathfrak{A}_2 \;=\; \sigma^*\overline{(\tau^*\mathfrak{A}_3)} \;\subseteq\; \overline{\sigma^*(\tau^*\mathfrak{A}_3)} \;=\; \overline{(\tau \circ \sigma)^*\mathfrak{A}_3} \;\subseteq\; \overline{\mathfrak{A}_1} \;\equiv\; \mathfrak{A}_1$$

implies that $(\tau \circ \sigma)^*\mathfrak{A}_3$ is dense in $\mathfrak{A}_1$.

### 2.4. Group Actions

This functoriality immediately allows us to lift group actions. For this, let $\mathcal{X}$ be some set and $\mathfrak{A}$ be again some $C^*$-subalgebra of $\ell^\infty(\mathcal{X})$. Recall that a (left) action of a group $S$ on $\mathcal{X}$ consists of mappings $\varphi_s : \mathcal{X} \longrightarrow \mathcal{X}$ that fulfill $\varphi_\mathbf{1} = \mathrm{id}$ as well as $\varphi_{s_1 s_2} = \varphi_{s_1} \circ \varphi_{s_2}$ for all $s_1, s_2 \in S$. Ignoring continuity or smoothness issues for the moment, we see that $\varphi_s$ can be lifted to the quantum level iff $\varphi_s^*\mathfrak{A} \subseteq \mathfrak{A}$. In order to lift all $\varphi_s$, we just have to require that $\mathfrak{A}$ is $S$-invariant, i.e., $\varphi_s^*\mathfrak{A} = \mathfrak{A}$ for all $s \in S$. Moreover, functoriality shows that $\overline{\varphi_{s_1 s_2}} = \overline{\varphi_{s_1}} \circ \overline{\varphi_{s_2}}$. As obviously the identity is lifted to the identity, we even see that each $\overline{\varphi_s}$ is a homeomorphism on $\overline{\mathcal{X}}$ with $\overline{\varphi_s}^{-1} = \overline{\varphi_{s^{-1}}}$. Altogether, we have

**Theorem 8.** *A group action on $\mathcal{X}$ lifts to a group action on $\overline{\mathcal{X}}$ iff $\mathfrak{A}$ is $S$-invariant. The lifted group action is even unique, and each group element acts by homeomorphisms on $\overline{\mathcal{X}}$ fulfilling*

$$\overline{\varphi}_s(\overline{x}) \;=\; \overline{x} \circ \varphi_s^* \tag{2}$$

*with $\overline{\varphi}$ denoting the lifted action of $S$ on $\overline{\mathcal{X}}$ defined by $\overline{\varphi}_s := \overline{\varphi_s}$.*

The issue of continuity is a bit more subtle. For topological groups $S$, one can show [39, 38] by methods from $C^*$-dynamical systems that $\overline{\varphi} : S \times \overline{\mathcal{X}} \longrightarrow \overline{\mathcal{X}}$ is continuous if $s \longmapsto a \circ \varphi_s$ is continuous for all $a \in \mathfrak{A}$; the converse is true if $\mathfrak{A}$ is unital. Remarkably, this is completely independent from whether $\varphi$ is continuous or not. Indeed, in the relevant cases of loop quantum cosmology or loop quantum gravity the action on the classical level is not necessarily smooth. But even in these cases, we already get for free that $\overline{\varphi}_s$ is always a homeomorphism.

### 2.5. Symmetric Sectors

The presence of a group action immediately leads us to the notion of invariant (or symmetric) elements. Here, the situation is even more interesting: we have group actions both on the classical and on the quantum level. Can we relate both levels?

To be more specific, first define the symmetric classical set by

$$\mathcal{X}_{\mathrm{red}} \; := \; \{x \in \mathcal{X} \mid \varphi_s(x) = x \quad \forall s \in S\}$$

and, analogously, the symmetric quantum space by

$$\overline{\mathcal{X}}_{\mathrm{red}} \; := \; \{\overline{x} \in \overline{\mathcal{X}} \mid \overline{\varphi}_s(\overline{x}) = \overline{x} \quad \forall s \in S\}.$$

At the same time, however, we can consider the space $\mathcal{X}_{\mathrm{red}}$ of symmetric classical configurations as a configuration space in its own right. This has been precisely the standard way of thinking in loop quantum cosmology. In particular, we can quantize it given some reasonable abelian $C^*$-algebra $\mathfrak{A}_{\mathrm{red}}$ of bounded functions on it. This algebra can, in principle, be chosen quite arbitrarily. If, however, one wants to embed the quantized reduced space $\overline{\mathcal{X}_{\mathrm{red}}}$ into the quantized full space $\overline{\mathcal{X}}$, one has to define $\mathfrak{A}_{\mathrm{red}}$ by $\overline{\mathfrak{A}|_{\mathcal{X}_{\mathrm{red}}}}$. It is now natural to ask how the spaces $\overline{\mathcal{X}}_{\mathrm{red}}$ and $\overline{\mathcal{X}_{\mathrm{red}}}$ as well as the closure of $\iota(\mathcal{X}_{\mathrm{red}})$ in $\overline{\mathcal{X}}$ are related. Here, $\mathcal{X}_{\mathrm{red}}$ is considered as a subset of $\mathcal{X}$.

**Theorem 9.** *We have*
1. *$\overline{\mathcal{X}_{\mathrm{red}}}$ is homeomorphic to $\overline{\iota(\mathcal{X}_{\mathrm{red}})}$, provided $\mathfrak{A}$ is unital.*
2. *$\overline{\iota(\mathcal{X}_{\mathrm{red}})}$ is contained in $\overline{\mathcal{X}}_{\mathrm{red}}$.*
3. *$\overline{\mathcal{X}}_{\mathrm{red}}$ is closed in $\overline{\mathcal{X}}$.*

Note that later we will identify $\overline{\mathcal{X}_{\mathrm{red}}}$ and $\overline{\iota(\mathcal{X}_{\mathrm{red}})}$.

*Proof.*

1. Let $\sigma : \mathcal{X}_{\mathrm{red}} \longrightarrow \mathcal{X}$ be the inclusion mapping[6] and $\overline{\sigma}$ the corresponding lift:

$$\begin{array}{ccc} \mathcal{X}_{\mathrm{red}} & \xrightarrow{\sigma} & \mathcal{X} \\ \downarrow{\iota_{\mathrm{red}}} & & \downarrow{\iota} \\ \overline{\mathcal{X}_{\mathrm{red}}} & \xrightarrow{\overline{\sigma}} & \overline{\mathcal{X}} \end{array}$$

Now, $\overline{\sigma} \circ \iota_{\mathrm{red}} = \iota \circ \sigma$ and the continuity of $\overline{\sigma}$ give

$$\begin{aligned} \iota(\sigma(\mathcal{X}_{\mathrm{red}})) & \equiv \overline{\sigma}(\iota_{\mathrm{red}}(\mathcal{X}_{\mathrm{red}})) \\ & \subseteq \overline{\sigma}(\overline{\iota_{\mathrm{red}}(\mathcal{X}_{\mathrm{red}})}) \subseteq \overline{\overline{\sigma}(\iota_{\mathrm{red}}(\mathcal{X}_{\mathrm{red}}))} \equiv \overline{\iota(\sigma(\mathcal{X}_{\mathrm{red}}))}, \end{aligned}$$

hence, by compactness of $\overline{\iota_{\mathrm{red}}(\mathcal{X}_{\mathrm{red}})}$,

$$\overline{\iota(\mathcal{X}_{\mathrm{red}})} \equiv \overline{\iota(\sigma(\mathcal{X}_{\mathrm{red}}))} = \overline{\sigma}(\overline{\iota_{\mathrm{red}}(\mathcal{X}_{\mathrm{red}})}) \equiv \overline{\sigma}(\overline{\mathcal{X}_{\mathrm{red}}}).$$

---

[6]Note that later we will refrain from writing $\sigma$ in the case of subspaces of invariant elements.

Now, the proof follows as $\overline{\sigma}$, by unitality, maps a compact space into a Hausdorff space, hence is a homeomorphism onto its image.

2. Since $\overline{\varphi}_s \circ \iota = \iota \circ \varphi_s$, each element of $\iota(\mathcal{X}_{\text{red}})$ is invariant under each $\overline{\varphi}_s$.
3. If the net $(\overline{x}_\alpha) \subseteq \overline{\mathcal{X}}$ converges to $\overline{x} \in \overline{\mathcal{X}}$, then $\overline{\varphi}_s(\overline{x}_\alpha) \to \overline{\varphi}_s(\overline{x})$ by continuity of $\overline{\varphi}_s$. Closedness of $\overline{\mathcal{X}}_{\text{red}}$ is now obvious. □

It now remains to decide whether the inclusion relation between $\overline{\mathcal{X}_{\text{red}}}$ and $\overline{\mathcal{X}}_{\text{red}}$ is proper. More literally: can any invariant quantum configuration be approximated by invariant classical configurations? Seen functorially on the level of configuration spaces, this means: do quantization and symmetry implementation commute?

The general answer is – well, it depends. Indeed, we are neither aware of nor have we been able to so far find general conditions that imply commutativity – or non-commutativity. In loop quantum cosmology, however, the usual answer is negative: quantization and reduction do not commute. We will discuss this issue in Subsection 4.8.

## 3. Loop Quantum Gravity

Having settled the general mathematical framework for the quantization scheme used in loop quantization, we are now going to apply it to gravity. First, in Subsection 3.1, we review the classical configuration space of general relativity, both for metric and for Ashtekar variables. Then, following [9, 25], we describe its loop quantum counterpart in three equivalent forms: as the spectrum of a $C^*$-algebra, as homomorphisms of the path groupoid and as a projective limit of Lie groups. The Ashtekar-Lewandowski measure [10] is described in Subsection 3.5, before we conclude with an explicit example [25] of an element of the quantum configuration space and with the discussion of the underlying smoothness category.

### 3.1. Classical Configuration Space

As we have learned in Section 2, we need two pieces of information to define the quantum configuration space of a physical theory: the classical configuration space $\mathcal{X}$ and some algebra $\mathfrak{A}$ of bounded functions thereon. Let us start with the classical configuration space, now for gravity.

**3.1.1. Metric Variables.** Classical gravity, or general relativity, is described by a spacetime metric $g$ fulfilling the Einstein equations $G + \Lambda g = 8\pi T$. Here, a spacetime metric is a metric of Lorentzian signature on a 4-dimensional, connected, time-oriented and oriented manifold. Moreover, the Einstein tensor $G$ is defined by $G := R - \frac{1}{2} R_{\text{scal}} g$, with $R$ and $R_{\text{scal}}$ being the Ricci and the scalar curvatures of $g$, respectively. Finally, $\Lambda$ is the cosmological constant and $T$ is the stress-energy tensor. We will set both terms to zero, this way restricting ourselves to matter-free gravity in the vacuum.

Conceptually, general relativity is a covariant theory. Technically, it requires to solve a nonlinear partial differential equation. At the same time, it is desirable to formulate general relativity as an initial value problem, i.e. as a canonical theory.

However, canonical theories require some distinguished time variable, not naturally given in relativistic theories. Nevertheless, there is a class of spacetimes that admit a well-posed initial value formulation, namely globally hyperbolic spacetimes [54]. Originally, this notion had referred to the causal structure of the spacetime metric. According to the celebrated results by Bernal and Sánchez [17, 16], it is equivalent to the assumption that the spacetime manifold $M$ is diffeomorphic to $\mathbb{R} \times \Sigma$ for some 3-manifold $\Sigma$ and that the Lorentzian metric is given by $-f_\tau \, \mathrm{d}\tau^2 + q_\tau$ for some smooth families $f_\tau$ and $q_\tau$ of functions and Riemannian metrics on $\Sigma$, respectively. Moreover, each level set $\Sigma_\tau := \{\tau\} \times \Sigma$ is then a Cauchy surface; usually, one simply identifies $\Sigma$ with $\Sigma_0$.

Now, given global hyperbolicity, we can consider Einstein's equations as an initial-value problem. For this, one first observes that any spacetime metric on $M$ can be uniquely reconstructed from its restriction $q$ to $\Sigma_\tau$ and from the second fundamental form (or, extrinsic curvature) $K$ describing the shape of $\Sigma_\tau$ as a submanifold of $M$. Einstein's equations transfer to some evolution equations for $q$ and $K$, but also imply that these two objects have to fulfill certain constraint equations, namely the diffeomorphism constraint and the Hamilton constraint. They encode the invariance of the theory under spatial and temporal diffeomorphisms. It is a very remarkable result that our initial-value problem can now be solved at least locally [34, 50, 48] iff these constraints are met. One can even show that the metric $q$ on $\Sigma$ and some expression being linear (up to some weight) in $K$ are canonically conjugate variables. Thus, we can consider the space of Riemannian 3-metrics on $\Sigma$ as the (unconstrained) configuration space of canonical general relativity.[7]

**3.1.2. Ashtekar Variables.** The configuration space of Riemannian metrics became the starting point of the Wheeler-DeWitt approach to quantum gravity. Its idea was basically to consider quantum gravity as quantum mechanics of metrics, with the Hamilton constraint inducing the Hamiltonian. Technically, the non-polynomiality of the Hamilton constraint, however, was a big issue. In the 1980s, Ashtekar [3, 2] tried to cure that problem introducing new variables, now carrying his name.

First, he encoded the metric by means of triads. Of course, each triad determines a metric by requiring it to be oriented orthonormal. However, the other direction is not unique. In fact, each metric admits a whole bunch of oriented orthonormal triads; more precisely, two triads give the same metric iff they are related by some $SO(3)$ element.

Second, he introduced a covariant derivative (or, Ashtekar connection) $\nabla^A$ on $\Sigma$. As triads comprise the information on the metric, the covariant derivative contains that on the second fundamental form. For this, recall that the latter one is the quadratic form defined by the Weingarten mapping $W$ with $W(X) := {}^4\nabla_X n$.

---

[7] Alternatively, one can also consider the space of all Riemannian 3-metrics modulo diffeomorphisms. This leads to the so-called superspace. However, its mathematical structure is rather complicated. [36]

Here, $n$ is the normal to $\Sigma$ within $(M, g)$, and $^4\nabla$ is the Levi-Civita connection on $(M, g)$. Now, the Ashtekar connection is defined by [33, 26]

$$\nabla^A_X Y := \nabla_X Y + \beta\, W(X) \times Y\,.$$

Here, the vector product has been transferred from $\mathbb{R}^3$ to $T\Sigma$ using a $q$-orthonormal triad. The complex number $\beta$ is a parameter [15, 42] of the theory and is called Barbero-Immirzi parameter.

Ashtekar observed that the connection and the triad form a canonical pair of variables. At the same time, there appeared an additional constraint, the so-called Gauß constraint that originates in the gauge freedom introduced via replacing metrics by triads. This, however, is no big deal as now, having connections as variables, we are in the realm of gauge field theories. There, gauge invariance as incorporated by the Gauß constraint can be tamed rather easily. The big advantage of the original Ashtekar variables became apparent, when the Hamilton constraint was calculated: for $\beta = \pm \mathrm{i}$, it is polynomial. This triggered the hope for a successful and rigorous canonical quantization of pure general relativity. However, as physical theories shall produce real numbers rather than complex ones, the choice of $\beta = \pm \mathrm{i}$ forced Ashtekar to introduce quite complicated reality conditions. It took about ten years until Thiemann [51, 53] realized that the non-polynomiality which had remained for $\beta \neq \pm \mathrm{i}$, can be expressed by means of the Poisson bracket and some appropriate volume function.

For our purposes, however, these considerations are less relevant. We are going to study the configuration space only – neither the phase space nor the constraints. Thus, the main lesson we should keep in mind, is: general relativity, using Ashtekar's variables, can be formulated as a gauge field theory[8], i.e., its configuration space is spanned by the covariant derivatives on $T\Sigma$. Alternatively, one can consider $T\Sigma$ as an vector bundle that is associated to any[9] spin bundle over $\Sigma$. The latter one is always an $SU(2)$ principal fibre bundle over $\Sigma$. Its connections are in one-to-one correspondence with covariant derivatives on the tangent bundle over $\Sigma$. – To summarize:

**Definition 10.** *Let $G$ be a connected compact Lie group, $M$ a manifold, and $P$ a $G$-principal fibre bundle over $M$. Then $\mathcal{A}$ denotes the set of all connections in $P$.*

Now, the **classical configuration space** of canonical general relativity is the set of all connections in an $SU(2)$ principal fibre bundle over some 3-dimensional manifold. Until the end of this section, however, we will not restrict ourselves to $G = SU(2)$ and $\dim M = 3$, but use $\mathcal{A}$ is the broader sense of the definition above.[10]

---

[8]Note, however, that for $\beta \neq \pm \mathrm{i}$, it does no longer correspond to a space-time gauge field theory, as the Ashtekar connection cannot be extended to a full space-time connection. [49]

[9]There are spin structures, hence spin bundles, as $\Sigma$ is orientable. [43]

[10]Note that neither for gravity nor for cosmology we will discuss the issue of gauge transforms. Indeed, the space of connections modulo gauge transforms usually serves as configuration space of gauge theories. A careful implementation of gauge transforms into our framework is still lacking, although we do not expect severe problems. Nevertheless, the presentation would probably be much more technical, whence we would have focused just on spaces of connections anyway.

### 3.2. Quantum Configuration Space

Having settled the classical configuration space $\mathcal{A}$ for general relativity, we now have to select an appropriate algebra of functions on $\mathcal{A}$. Here, the motivation comes from lattice gauge field theories. There, rather than connections themselves, one considers parallel transports along the bonds of a lattice that is typically cubic. Here, we will now consider parallel transports just along arbitrary paths. Indeed, one can reconstruct a connection as soon as one knows all of its parallel transports.

Using all parallel transports, however, is technically a bit more subtle than in usual lattice gauge field theory. There, one always (at least implicitly) assumes to be given some (possibly global) trivialization of the bundle. This allows parallel transports to be just elements of the structure group. Actually, parallel transports are mappings $h_{\gamma,A} : P_{\gamma(0)} \longrightarrow P_{\gamma(1)}$ from the fibre over the starting point to the fibre over the end point of the path $\gamma$ under consideration. Of course, each fibre is (non-canonically) isomorphic to the structure group $G$; for instance, choosing $\nu(m)$ in the fibre $P_m$ over $m$, we get such an isomorphism $G \longrightarrow P_m$ via $g \longmapsto \nu(m)g$. Only now, as the mappings $h_{\gamma,A}$ commute with the right action of the structure group, one can identify them with elements of $G$ via $h_{\gamma,A}(\nu(\gamma(0))) = \nu(\gamma(1))h_\gamma^s(A)$. Here, the set-theoretic global section[11] $\nu : M \longrightarrow P$ is any function with $\pi \circ \nu = \mathrm{id}_M$ for $\pi : P \longrightarrow M$ the canonical projection.

Considering parallel transports as elements in $G$ rather than equivariant mappings between fibres, indeed, is much more convenient for our purposes. In fact, as $G$ is assumed compact, $h_\gamma^\nu$ can now be considered as a mapping from $\mathcal{A}$ to $G$, whose matrix element functions are bounded functions on $\mathcal{A}$ – exactly the structures we need for our algebra.

**Definition 11.** *Fix some set-theoretic global section in $P$ and let $\mathfrak{B}$ contain precisely the functions $f \circ h_\gamma^\nu : \mathcal{A} \longrightarrow \mathbb{C}$, with $f : G \longrightarrow \mathbb{C}$ running over the matrix functions[12] and $\gamma$ running over all analytic paths in $M$.*

*Then $\mathfrak{A}$ is defined to be the unital $C^*$-subalgebra of $\ell^\infty(\mathcal{A})$ generated by $\mathfrak{B}$.*

In the loop quantum gravity literature, $\mathfrak{B}$ (or, often, some slightly larger subsets of $\mathfrak{A}$) is usually referred to as the set of cylindrical functions. Note that $\mathfrak{A}$ does not depend on the choice of the set-theoretic section $\nu$. In fact, if $\mu$ is another section, there are some $g_x, g_y \in G$, such that $h_\gamma^\nu(A) = g_y^{-1} h_\gamma^\mu(A) g_x$ for all $A$ and all $\gamma$ connecting $x$ and $y$. Now, it is clear that we get the same $\mathfrak{A}$ in both situations. In the following, we will tacitly assume to have fixed some set-theoretic section for calculating parallel transport functions, and simply write $h_\gamma$ instead of $h_\gamma^\nu$.

According to Section 2, now the **quantum configuration space** of canonical general relativity in the loop approach is the spectrum of $\mathfrak{A}$. This is a compact Hausdorff space and denoted by $\overline{\mathcal{A}}$. Its elements are called generalized connections. By Proposition $1_2$ it contains $\mathcal{A}$ as a dense subspace.

---

[11] Note that we do not require $\nu$ to even be a local section in the bundle sense. In fact, $\nu$ need not be continuous; this is referred to by noting "set-theoretic". [29]

[12] We assume to have fixedly chosen $G$ as a Lie subgroup of some $U(n)$.

### 3.3. Groupoid Homomorphisms

At a first glance, spectra of abelian $C^*$-algebras need not look easily accessible. This, however, is not true for the set $\overline{\mathcal{A}}$ of generalized connections.

For this, let us go back to the case of "classical" connections or better their parallel transports. So far, we have always considered them as functions from $\mathcal{A}$ to $G$, indexed by the paths in $M$. Now, we will consider them as mappings $h_A : \mathcal{P} \longrightarrow G$ from the set $\mathcal{P}$ of paths to $G$ indexed by the connections. Of course, $h_A$ will be smooth in a certain sense, but that is not the property most relevant for us. Instead, $h_A$ is multiplicative. In fact, given a path $\gamma$ cut in two pieces $\gamma_1$ and $\gamma_2$, the parallel transports obey $h_A(\gamma) \equiv h_A(\gamma_1 \gamma_2) = h_A(\gamma_1) h_A(\gamma_2)$. Similarly, $h_A(\gamma^{-1}) = h_A(\gamma)^{-1}$, for $\gamma^{-1}$ being the path $\gamma$ passed in the opposite direction. Thus, giving the set $\mathcal{P}$ of analytic paths the structure of a groupoid[13], each $h_A$ is a homomorphism from $\mathcal{P}$ to $G$. And this is the key to the spectrum, giving

**Proposition 12.** $\overline{\mathcal{A}}$ *is isomorphic to* $\mathrm{Hom}(\mathcal{P}, G)$.

This means, $\overline{\mathcal{A}}$ consists of *all* homomorphisms, not just the sort-of-smooth ones. The isomorphism itself is rather simply described. We may extend $\overline{A} \in \overline{\mathcal{A}}$ to

$$\overline{A} \;:\; \mathfrak{A} \otimes \mathbb{C}^{n \times n} \;\longrightarrow\; \mathbb{C}^{n \times n},$$

with our compact $G$ embedded into $U(n) \subseteq \mathbb{C}^{n \times n}$. Then we just have

$$\overline{A}(h_\gamma) \;=\; \overline{A}(\gamma),$$

where the right-hand $\overline{A} : \mathcal{P} \longrightarrow G$ is the homomorphism that corresponds to the left-hand $\overline{A} \in \overline{\mathcal{A}}$ (extended as above). That the mapping above is indeed an isomorphism follows from the homomorphy of parallel transports, and as $h_\gamma$ equals $h_\delta$ iff $\gamma$ and $\delta$ coincide up to the parametrization.

We will identify $\overline{\mathcal{A}}$ and $\mathrm{Hom}(\mathcal{P}, G)$ in the following.

### 3.4. Projective Limits

Beyond the formulations of $\overline{\mathcal{A}}$ via $C^*$-algebras and via path groupoids, there is a third one using projective limits. For this, consider any analytic graph $\boldsymbol{\gamma}$, i.e., a finite graph whose edges are analytic paths $\gamma_i$. Any such $\boldsymbol{\gamma}$ defines a surjective continuous mapping $h_{\boldsymbol{\gamma}} : \overline{\mathcal{A}} \longrightarrow G^n$ via $h_{\boldsymbol{\gamma}}(\overline{A}) = (\overline{A}(\gamma_1), \ldots, \overline{A}(\gamma_n))$. These mappings $h_{\boldsymbol{\gamma}}$ can be seen as canonical projections of an appropriate projective limit.

Indeed, we define an ordering on the set of analytic paths as follows. We say $\boldsymbol{\gamma} \leq \boldsymbol{\delta}$ iff any edge in $\boldsymbol{\gamma}$ can be decomposed into edges of $\boldsymbol{\delta}$ or their inverses. Corresponding to these decompositions, we define a projection

$$\pi_{\boldsymbol{\gamma}}^{\boldsymbol{\delta}} \;:\; G^{\#\boldsymbol{\delta}} \longrightarrow G^{\#\boldsymbol{\gamma}}.$$

For instance, if $\gamma_1 = \delta_1 \delta_2$ and $\gamma_2 = \delta_3^{-1}$, then $\pi_{\boldsymbol{\gamma}}^{\boldsymbol{\delta}}(g_1, g_2, g_3) = (g_1 g_2, g_3^{-1})$. As

$$\pi_{\boldsymbol{\gamma}}^{\boldsymbol{\delta}} \circ \pi_{\boldsymbol{\delta}}^{\boldsymbol{\varepsilon}} \;=\; \pi_{\boldsymbol{\gamma}}^{\boldsymbol{\varepsilon}} \qquad \text{for } \boldsymbol{\gamma} \leq \boldsymbol{\delta} \leq \boldsymbol{\varepsilon},$$

these mappings form a projective system. One can easily show that

---

[13] For this, one has to identify paths that coincide up to their parametrization.

**Proposition 13.** $\overline{\mathcal{A}}$ *is homeomorphic to* $\varprojlim_{\gamma} G^{\gamma}$ *via* $\overline{A} \longmapsto (h_{\gamma}(\overline{A}))_{\gamma}$.

In particular, the canonical projections of the projective limit are just the $h_{\gamma}$.

### 3.5. Ashtekar-Lewandowski Measure

One of the main breakthroughs of loop quantum gravity was not only the definition of a compact quantum configuration space for general relativity, but even more the rigorous construction of a normalized Radon measure thereon, namely the Ashtekar-Lewandowski measure. Its construction is very easy using the projective limit framework, given the underlying ordering is directed which is the case for analytic graphs. Recall that then the normalized Radon measures on a projective limit are in one-to-one correspondence to consistent families of normalized Radon measures on the constituents of the limit. In our case, this means that the measures on $\overline{\mathcal{A}}$ correspond to families $(\mu_{\gamma})_{\gamma}$ of measures on the $G^{\#\gamma}$ that fulfill

$$\mu_{\boldsymbol{\gamma}} \;=\; (\pi^{\boldsymbol{\delta}}_{\boldsymbol{\gamma}})_{*}\mu_{\boldsymbol{\delta}} \qquad \text{for all } \boldsymbol{\gamma} \leq \boldsymbol{\delta}.$$

The most obvious choice is to let each $\mu_{\gamma}$ be the normalized Haar measure. Indeed, this family is consistent and corresponds to the Ashtekar-Lewandowski measure $\mu_{\text{AL}}$. This measure underlies all the constructions in loop quantum gravity. In particular, it provides us with a kinematical Hilbert space $L^2(\overline{\mathcal{A}}, \mu_{\text{AL}})$.

Of course, there are many more measures on $\overline{\mathcal{A}}$, but the choice of the Ashtekar-Lewandowski is indeed natural. There are strong uniqueness results that single out $\mu_{\text{AL}}$. In particular, up to some technical assumptions like regularity and cyclicity, both the holonomy-flux algebra [44] and the Weyl algebra [30] can be represented only on $L^2(\overline{\mathcal{A}}, \mu_{\text{AL}})$ in a diffeomorphism-covariant way.

### 3.6. Explicit Example

Having laid out the theory, we should show that we have made a nontrivial extension of the classical configuration space. So let us explicitly construct generalized, but non-smooth connections. For this, fix some path $\delta$ as well as some $g \in G$, and let $n(\gamma)$ tell us how often $\gamma$ passes an initial segment of $\delta$ (counted negatively if passed in the other direction). Now, $\overline{A}_g(\gamma) := g^{n(\gamma)}$ is a well-defined generalized connection. If, for simplicity, we assume $P$ to be trivial, then it is a smooth connection iff $g = \mathbf{1}$, showing that $\mathcal{A}$ is properly contained in $\overline{\mathcal{A}}$. For general $P$, the argumentation is similar.

### 3.7. Paths

When defining $\overline{\mathcal{A}}$ to be the quantum configuration space, we have assumed without further comment that the algebra $\mathfrak{A}$ is generated by the parallel transport functions along all analytic paths in $M$. Why have we done this? Why have we not used further paths? Why have we not restricted ourselves to smaller classes?

Well, first of all, analytic objects behave very nicely. Indeed, if two analytic paths have infinitely many intersection points, they have to coincide on full intervals which may even be taken maximal. This implies that two analytic graphs are always subgraphs of a third analytic graph, i.e., the ordering on analytic graphs introduced above is direct. This has turned out crucial for the construction of measures.

Of course, one has to grant that one can weaken the smoothness requirements the paths are to fulfill. For that purpose, webs [14, 13, 27, 28] in the smooth category or hyphs [23] in the arbitrary $C^k$ categories have been introduced and widely studied. Indeed, significant parts of the theory have been transferred. Nevertheless, the use of paths that are non-analytic, has lead to enormous technical difficulties. Thus, it is at least a matter of convenience to choose the analytic category for all the manifolds, bundles, and paths in the game.

Conceptually, there is even a further justification for choosing analytic paths. Actually, general relativity seems to contradict analyticity, as the former one is a local, the latter one a non-local structure. However, this puzzle can be resolved using semianalytic structures [45, 40, 30, 44] rather than analytic ones. Indeed, semianalyticity provides us with non-local, i.e. non-rigid structures. Even more, the loop quantum gravity uniqueness theorems mentioned above work in the semianalytic framework. For paths, semianalytic just means piecewise analytic, which is the same as to admit the concatenation of analytic paths. At the same time, the $C^*$-algebra $\mathfrak{A}$ generated by the analytic paths is the same as those generated by piecewise analytic paths. Therefore, at the level of graphs it is completely justified to restrict oneself to analytic paths.

It remains to discuss whether we might further restrict the set of paths, say to straight lines when the base manifold is $\mathbb{R}^3$. Indeed, there is no obstruction to do this. One just has to guarantee that with any admissible path all of its subpaths as well as its inverse are admissible. These issues will become relevant in cosmological models and will be discussed in more detail in Subsection 4.5.

## 4. Homogeneous Isotropic Loop Quantum Cosmology

Now we are going to apply our results to the case of $k = 0$ homogeneous isotropic cosmology. After recalling the implications of homogeneity and isotropy as well as explaining $k$, we will state the space of classical invariant connections. Then we will investigate the algebra of functions whose spectrum can be continuously embedded into the full space of generalized connections, in the analytic case. We also discuss other smoothness categories focusing on their embedding properties. Afterwards, we review the invariance of measures and close with the noncommutativity of symmetry reduction and quantization.

### 4.1. Euclidean Group Action

In general, homogeneity and isotropy imply that the sectional curvature $k$ of the metric on the Cauchy surface $\Sigma$ is constant. [54] If $k$ even vanishes, then $\Sigma$ is necessarily covered by $\mathbb{R}^3$. As we will assume that the Cauchy surface is even simply connected, $\Sigma$ equals the Euclidean $\mathbb{R}^3$. Homogeneity and isotropy now mean that our theory shall be invariant under all rigid motions, hence w.r.t. the connected component $E$ of the Euclidean group in 3 dimensions.

As we are going to study gravity and cosmology in their formulations using Ashtekar variables, we have to consider now connections in an $SU(2)$-principal fibre bundle $P$ over $\Sigma = \mathbb{R}^3$. As $\Sigma$ is contractible, the bundle is trivial, hence $P = \Sigma \times SU(2)$. The delicate point is how to lift the action of the Euclidean group from $\Sigma$ to $P$ and to the connections in $P$. The action on the $\Sigma$-part, of course, remains to be the defining action of the Euclidean group $E$. But, how to act on the $SU(2)$-part? Here, one should assume that not only $E$ is acting, but even its covering group, which is homeomorphic to $\mathbb{R}^3 \times S^3$. Finally, this leads to the following situation [38]:

**Definition 14.** *The* **symmetry group** *of $k=0$ homogeneous isotropic cosmology is*

$$S \;=\; \mathbb{R}^3 \rtimes SU(2)\,.$$

*It acts from the left on $P = \mathbb{R}^3 \times SU(2)$ via*

$$(x,r) \circ (y,g) \;\longmapsto\; (ry + x, rg)\,.$$

Here, $SU(2)$ acts on $\mathbb{R}^3$ via first projecting $SU(2)$ to $SO(3)$ by the covering map. More explicitly, $[ry]^i \tau_i = \mathrm{Ad}_r(y^j \tau_j)$ for any $\mathfrak{su}(2)$ basis $\{\tau_i\}$ with $[\tau_i, \tau_j] = \varepsilon_{ij}^m \tau_m$.

### 4.2. Invariant Connections

In the next step we have to identify the elements of $\mathcal{A}_{\mathrm{red}}$, i.e. the connections in $P$ that are invariant under the action of the symmetry group $S$ on $P$.

**Proposition 15.** $\mathcal{A}_{\mathrm{red}}$ *is isomorphic to $\mathbb{R}$. More precisely,*

$$\mathcal{A}_{\mathrm{red}} \;=\; \{\theta + \mathbf{c}\,\eta \mid \mathbf{c} \in \mathbb{R}\}$$

*with $\theta$ being the Maurer-Cartan form on $G = SU(2)$ and $\eta \in \Omega^1(P, \mathfrak{su}(2))$ with*

$$\eta_{(x,g)} \;=\; \mathrm{Ad}_g \tau_i \otimes \mathrm{d}x^i \qquad \textit{for all } (x,g) \in P\,.$$

That the forms above are indeed invariant forms, can be calculated immediately. More elaborate is the proof that there are not any other invariant forms. [38]

The more common version of the forms above is their pullbacks to the base manifold using the canonical section into the trivial bundle $P = \mathbb{R}^3 \times SU(2)$. Indeed, these pullback forms are $\mathbf{c}\,\tau_i \otimes \mathrm{d}x^i$.

### 4.3. Quantizing Algebra

Recall that we are aiming at constructing a version of loop quantum cosmology whose configuration space can be embedded into that of full loop quantum gravity. Therefore, we have to define the cosmological cylindrical functions as the restrictions of the gravitational cylindrical functions (see Definition 11) to $\mathcal{A}_{\mathrm{red}} \subseteq \mathcal{A}$. For this, let us study first how parallel transports depend on $\mathcal{A}_{\mathrm{red}}$.

Let there be given a connection $A \in \mathcal{A}$. We identify $A$ with the corresponding connection one-form on $M$ that one gets via pullback along the standard section in $P$; recall that $P$ is trivial. Moreover, let $\gamma : [0,1] \longrightarrow M$ be some path in $M$ and let $\gamma_t$ be its restriction to $[0, t]$. Finally, denote by $\boldsymbol{g}(t) := h_A(\gamma_t)$ the parallel transport along $\gamma_t$. Then $\boldsymbol{g}$ obeys the differential equation

$$\dot{\boldsymbol{g}} = -A(\dot{\gamma})\boldsymbol{g} \qquad \text{with } \boldsymbol{g}(0) = \mathbf{1}. \tag{3}$$

Specializing to $\theta + \mathbf{c}\eta \in \mathcal{A}_{\text{red}} \cong \mathbb{R}$, any matrix function $a$ of $\boldsymbol{g}$ fulfills [21]

$$\ddot{a} + \mathbf{c}^2 a = \mathbf{c} f_0 a + f_1 \dot{a} \tag{4}$$

for certain $f_0, f_1 \in C[0, 1]$ that can be expressed using $\dot{\gamma}$; moreover, $a(0) \in \mathbb{C}$ and $\dot{a}(0) \in \mathbb{C} + \mathbf{c}\mathbb{C}$ depend on $\gamma$. One can show that $f_0$ and $f_1$ vanish iff $\gamma$ is a straight line; they are constant iff $\gamma$ is a spiral arc, comprising the cases of straight lines and circles. [21, 24]

In particular, we have now learned that the cosmological, i.e. reduced quantizing algebra $\mathfrak{A}_{\text{red}}$ is generated by the functions $\mathbf{c} \longmapsto a(t)$, whereas $a$ solves (4), $t$ runs over $[0, 1]$, and $\gamma$ (which determines coefficients and initial values for (4)) runs over all analytic paths in $\mathbb{R}^3$. Note that $a$, of course, depends on $\mathbf{c}$ via (4).

Let us calculate $\mathfrak{A}_{\text{red}}$. First consider straight lines. Here, (4) has the simple form $\ddot{a} + \mathbf{c}^2 a = 0$. Hence, $a(t)$ is a linear combination of $e^{it\mathbf{c}}$ and $e^{-it\mathbf{c}}$. Carefully handling the possible initial values, we see that any character $\mathbf{c} \longmapsto e^{it\mathbf{c}}$ is contained in $\mathfrak{A}_{\text{red}}$. If, next, we look at circles with radius $\frac{1}{2\lambda}$, (4) leads to $\ddot{a} + (\mathbf{c}^2 + \lambda^2)a = 0$ (after some $t$-dependent rescaling of $a$, irrelevant for our purposes). Its solutions $a$ can be written as a linear combination of exponentials of $\pm it\sqrt{\mathbf{c}^2 + \lambda^2}$. Focusing on their $\mathbf{c}$-dependence, we see that for large $\mathbf{c}$ they behave like periodic functions. A more careful investigation shows that $a(t)$ is indeed a linear combination of a character of $\mathbb{R}$ and a function vanishing at infinity. Again, exhausting all $\lambda > 0$ and handling the initial values properly, we see that we can even approximate any sum of this type. This shows that $\mathfrak{A}_{\text{red}}$ at least contains $C_{\text{AP}}(\mathbb{R}) + C_0(\mathbb{R})$, i.e. all sums of an almost periodic function and a function vanishing at infinity. It is not very difficult, but technically a bit more elaborate to show that these sums even comprise all possibilities: [24]

**Proposition 16.** *The **quantizing algebra** for loop quantum cosmology is*

$$\mathfrak{A}_{\text{red}} = C_0(\mathbb{R}) + C_{\text{AP}}(\mathbb{R}).$$

Note that $\mathfrak{A}_{\text{red}}$ is indeed a $C^*$-algebra [46, 33].

**4.4. Generalized Invariant Connections**

Our next task is to determine the spectrum $\overline{\mathbb{R}} \equiv \overline{\mathcal{A}_{\text{red}}}$ of $\mathfrak{A}_{\text{red}}$. [31] If the sum of $C_0(\mathbb{R})$ and $C_{\text{AP}}(\mathbb{R})$ was direct in the sense of $C^*$-algebras, this would be an easy game. However, obviously, this is not the case; the sum is only direct in the sense of vector spaces. Nevertheless, there is a bit beyond nothing. Indeed, observe that both algebras are $C^*$-subalgebras of $\ell^\infty(\mathbb{R})$ with $C_0(\mathbb{R})$ even being an ideal. This already suffices to show that the spectrum of $\mathfrak{A}_{\text{red}}$ *as a set* equals the disjoint sum of the two single spectra, namely, $\mathbb{R}$ and the Bohr compactification $\mathbb{R}_{\text{Bohr}}$, respectively. To see this, define

$$\tau(\mathbf{c}) \;:\; a_0 + a_1 \longmapsto a_0(\mathbf{c}) + a_1(\mathbf{c}) \tag{5}$$

$$\tau(\mathbf{b}) \;:\; a_0 + a_1 \longmapsto \quad\quad \mathbf{b}(a_1) \tag{6}$$

for $\mathbf{c} \in \mathbb{R}$, $\mathbf{b} \in \mathbb{R}_{\text{Bohr}} \equiv \operatorname{spec} C_{\text{AP}}(\mathbb{R})$, $a_0 + a_1 \in C_0(\mathbb{R}) + C_{\text{AP}}(\mathbb{R})$. From

$$\begin{aligned}
[\tau(\mathbf{b})]\big((a_0 + a_1)(b_0 + b_1)\big) &= [\tau(\mathbf{b})]\big((a_0 b_0 + a_0 b_1 + a_1 b_0) + a_1 b_1\big) \\
&= \mathbf{b}(a_1 b_1) \;=\; \mathbf{b}(a_1)\,\mathbf{b}(b_1) \\
&= [\tau(\mathbf{b})](a_0 + a_1)\,[\tau(\mathbf{b})](b_0 + b_1)
\end{aligned}$$

we see that $\tau : \mathbb{R} \sqcup \mathbb{R}_{\text{Bohr}} \longrightarrow \overline{\mathbb{R}}$ is well defined. It is not very difficult to show that $\tau$ is even bijective, whence we will identify $\overline{\mathbb{R}}$ with $\mathbb{R} \sqcup \mathbb{R}_{\text{Bohr}}$.

The determination of the topology of $\mathbb{R}$, however, is more elaborate. [31] As the sum of $C_0(\mathbb{R})$ and $C_{\text{AP}}(\mathbb{R})$ is not direct, we cannot expect the topology to be the disjoint-sum topology. Indeed, it is not. Instead, it is twisted. It is generated by the following types of sets:

| | | |
|---|---|---|
| Type 1: | $V \sqcup \varnothing$ | with open $V \subseteq \mathbb{R}$ |
| Type 2: | $\complement K \sqcup \mathbb{R}_{\text{Bohr}}$ | with compact $K \subseteq \mathbb{R}$ |
| Type 3: | $a_1^{-1}(U) \sqcup \widetilde{a}_1^{-1}(U)$ | with open $U \subseteq \mathbb{C}$ and $a_1 \in C_{\text{AP}}(\mathbb{R})$. |

Nevertheless, the relative topologies on $\mathbb{R}$ and $\mathbb{R}_{\text{Bohr}}$ considered as subsets of $\overline{\mathbb{R}}$ are the usual ones. Also, $\mathbb{R}$ is open and dense in $\overline{\mathbb{R}}$. Moreover, the canonical mapping $\iota_{\text{red}} : \mathbb{R} \longrightarrow \overline{\mathbb{R}} \equiv \mathbb{R} \sqcup \mathbb{R}_{\text{Bohr}}$ is the identity on $\mathbb{R}$. Finally, as $\mathfrak{A}_{\text{red}}$ is unital, $\overline{\mathbb{R}}$ is compact as desired.

### 4.5. Embeddings

By construction, $\overline{\mathbb{R}}$ is continuously embedded into $\overline{\mathcal{A}}$ extending the natural embedding of $\mathbb{R} \equiv \mathcal{A}_{\text{red}}$ into $\mathcal{A}$. In fact, as we have defined $\mathfrak{A}_{\text{red}}$ to be generated by the parallel transport matrix functions restricted to $\mathcal{A}_{\text{red}}$, the embedding criterion of Corollary 7 is fulfilled automatically. However, there are also other versions of loop quantum cosmology that have been discussed in the literature. They differ from our strategy by using different quantizing algebras. Indeed, the original way Bojowald invented loop quantum cosmology, was to consider parallel transports along straight lines only. As we have seen above, these functions span the algebra of almost periodic functions whose spectrum is $\mathbb{R}_{\text{Bohr}}$. However, $C_{\text{AP}}(\mathbb{R})$ does *not* fulfill the embedding criterion of Corollary 7. In fact, we have seen that $\mathfrak{A}|_{\mathcal{A}_{\text{red}}}$ is *not* contained in $C_{\text{AP}}(\mathbb{R})$ although that is a necessary condition for embeddability.

Of course, these arguments are only valid assuming that the quantizing algebra for loop quantum gravity is indexed by all analytic graphs. Over the years, however, several other choices have been discussed: smooth paths [14, 13, 27], $C^k$ paths [23], piecewise linear paths [56, 22], paths in a fixed graph [35], or paths given by iterated barycentric subdivisions [1]. Similarly, in the cosmological arena piecewise linear paths [22, 19], parts of a fixed geodesic [19] or just two paths having incommensurable lengths [52] have been investigated. We have displayed in Table 1 for each combination whether the classical embedding $\mathcal{A}_{\text{red}} \subseteq \mathcal{A}$ does lift to the quantum level (+) or does not (−). Note that we have excluded the gravity cases based on a fixed graph or barycentric subdivisions, as there the quantizing algebras do, in general, not separate the classical configuration space [24], whence $\iota : \mathcal{A} \longrightarrow \overline{\mathcal{A}}$ is not injective there. It should now be no surprise that a continuous embedding is given if and only if the classes of paths used for the gravitational and for the cosmological theories coincide [24].

TABLE 1. Continuous Embeddings of Quantum Configuration Spaces

| paths for… gravity | …cosmology as for gravity | piecewise linear | in a geodesic | incommens'ble |
|---|---|---|---|---|
| piecewise analytic | + | − | − | − |
| piecewise smooth | + | − | − | − |
| piecewise $C^k$ | + | − | − | − |
| piecewise linear | + | + | + | − |

Consequently, the strategy presented above, i.e. defining the quantized algebras for gravity and cosmology using identical sets of paths is the only way to get embeddability. This, however, does not tell us a priori whether the choice for cosmology should follow that for gravity or vice versa. Indeed, in order to keep Bojowald's original straight-line cosmology, it has been argued [22] that one should restrict oneself to straight lines already at gravity level. We, however, think that this is not desirable as then other symmetries might not be treatable within full gravity. Therefore, we prefer our strategy "cosmology follows gravity".

### 4.6. Invariant Measures

In order to define a kinematical Hilbert space and to then obtain quantum states, one should look for finite measures on the respective quantum configuration space. For full gravity, the Ashtekar-Lewandowski measure $\mu_{\mathrm{AL}}$ on $\overline{\mathcal{A}}$ turned out to be the measure distinguished by strong uniqueness results [30, 44]. In Bojowald's cosmology, the Haar measure on $\mathbb{R}_{\mathrm{Bohr}}$ adopted that rôle [5]. What could now be the measure of choice in our version of loop quantum cosmology?

Well, the easiest way would be to choose a Haar measure on the quantum configuration space $\overline{\mathbb{R}} = \mathbb{R} \sqcup \mathbb{R}_{\mathrm{Bohr}}$. However, as Hanusch [39] pointed out, $\overline{\mathbb{R}}$ does not carry the structure of a topological group compatible with that on $\mathbb{R}$. No group structure – no Haar measure.

Thus, maybe we should lower our sights. Indeed, a Haar measure does not exploit the full structure of a topological group. Instead, it is defined to be invariant w.r.t. the left translation of the group on itself. For the Bohr compactification, we even need less than full invariance:

**Lemma 17.** *Any finite $\mathbb{R}$-invariant Radon measure on $\mathbb{R}_{\mathrm{Bohr}}$ is $\mathbb{R}_{\mathrm{Bohr}}$-invariant.*

*Proof.* Let $U$ be a neighbourhood in $\mathbb{R}_{\mathrm{Bohr}}$ of some compact $K$ that fulfills $\mu(U) < \mu(K) + \varepsilon$, and let $\mathbf{b} \in \mathbb{R}_{\mathrm{Bohr}}$. By denseness of $\mathbb{R}$ in the compact group $\mathbb{R}_{\mathrm{Bohr}}$, there is some $\mathbf{c} \in \mathbb{R}$ with $\mathbf{b} + K \subseteq \mathbf{c} + U$, whence by $\mathbb{R}$-invariance

$$\mu(\mathbf{b} + K) \ \leq \ \mu(\mathbf{c} + U) \ = \ \mu(U) \ < \ \mu(K) + \varepsilon\,.$$

This gives $\mu(\mathbf{b} + K) = \mu(K)$. Now the proof follows from regularity. □

On the other hand, using the general strategy from Subsection 2.4, the $\mathbb{R}$-action on $\mathbb{R}_{\mathrm{Bohr}}$ can be considered as the lift of the standard $\mathbb{R}$-action on $\mathbb{R}$. For this, just observe that $\mathbb{R}_{\mathrm{Bohr}}$ is nothing but the quantum configuration space for $\mathcal{X} = \mathbb{R}$ and quantizing algebra $C_{\mathrm{AP}}(\mathbb{R})$. But, why should we look for $\mathbb{R}$-invariant measures at all? Well, for the same reason as in full gravity. In fact, the Ashtekar-Lewandowski measure is invariant under all exponentiated fluxes. In cosmology, their counterpart is just translations by real numbers. Therefore, it is very natural to ask for $\mathbb{R}$-invariant measures in the cosmological case. The Haar measure, as shown above, is indeed the only candidate in Bojowald's construction of loop quantum cosmology; in fact, here the quantizing algebra has been $C_{\mathrm{AP}}(\mathbb{R})$. In our framework, however, the quantizing algebra is $\mathfrak{A}_{\mathrm{red}} = C_0(\mathbb{R}) + C_{\mathrm{AP}}(\mathbb{R})$. Do we get an $\mathbb{R}$-action there as well? Yes, of course. Theorem 8 tells us that the only condition is the $\mathbb{R}$-invariance of $\mathfrak{A}_{\mathrm{red}}$, which is obviously given. Even more, as the action preserves the direct sum $C_0(\mathbb{R}) + C_{\mathrm{AP}}(\mathbb{R})$, the restriction of the $\mathbb{R}$-action on $\overline{\mathbb{R}}$ to its components $\mathbb{R}$ and $\mathbb{R}_{\mathrm{Bohr}}$ coincides with the respective standard action. Finally, $\mathbf{x} \longmapsto L_{\mathbf{x}}^* a$ which maps $\mathbb{R}$ to $\mathfrak{A}_{\mathrm{red}}$, is continuous for all $a \in \mathfrak{A}_{\mathrm{red}}$ [38], whence the $\mathbb{R}$-action on $\overline{\mathbb{R}}$ is continuous as well by the closing remark in Subsection 2.4.

Now, we are left with just a final observation: although the topology of $\overline{\mathbb{R}} = \mathbb{R} \sqcup \mathbb{R}_{\mathrm{Bohr}}$ is not that of the disjoint sum, the Borel algebra of $\overline{\mathbb{R}}$ *is* the disjoint sum of the Borel algebras of $\mathbb{R}$ and of $\mathbb{R}_{\mathrm{Bohr}}$. This implies that the finite Radon measures on $\overline{\mathbb{R}}$ are the sums of finite Radon measures on $\mathbb{R}$ and on $\mathbb{R}_{\mathrm{Bohr}}$. [31, 38]

**Proposition 18.** *The Haar measure on the $\mathbb{R}_{\mathrm{Bohr}}$-part is the only normalized Radon measure on $\overline{\mathbb{R}}$ that is $\mathbb{R}$-invariant.* [38]

*Proof.* Let $\mu = \mu_0 + \mu_1$ be such a measure on $\overline{\mathbb{R}} = \mathbb{R} \sqcup \mathbb{R}_{\mathrm{Bohr}}$. Setting $I = [0, 1)$, we have $1 \geq \mu_0(\mathbb{R}) = \sum_{n \in \mathbb{Z}} \mu_0(n + I) = \sum_{\mathbb{Z}} \mu_0(I)$ by $\mathbb{R}$-invariance, which implies $\mu_0 = 0$. As now $\mu_1$ is $\mathbb{R}$-invariant, the proof follows from Lemma 17. □

This proposition has an astonishing consequence: it reconciles the original approach by Bojowald with that discussed here – both lead to the same kinematical Hilbert space. Our approach, however, has the advantage that it is functorial. This means that we can guarantee for a continuous embedding of the configuration spaces being a fundamental step in the Bojowald-Kastrup construction [20] of symmetric states in loop quantum gravity.

The bridge between loop quantum gravity and loop quantum cosmology that had been destroyed by the non-embedding results [21] of Brunnemann and the author, has now been rebuilt at another place.

**4.7. Invariant Generalized Connections**

So far, we have studied loop quantum cosmology as the quantum theory of the cosmological sector of gravity. In other words, we have first reduced the theory and then quantized. Does this give the same result as first quantizing and then reducing? In other words, is loop quantum cosmology the homogeneous isotropic sector of full loop quantum gravity?

From the general theory we know that we can lift the action of the classical symmetry group $S$ on $\mathcal{A}$ to an action on $\overline{\mathcal{A}}$. The generalized connections that are invariant w.r.t. the lifted action form the space $\overline{\mathcal{A}}_{\text{red}}$ which is a closed, hence compact subset of $\overline{\mathcal{A}}$. Considering $\overline{\mathcal{A}}_{\text{red}}$ as a subset of $\overline{\mathcal{A}}$, we already know that $\overline{\mathcal{A}}_{\text{red}}$ is a subset of $\overline{\mathcal{A}}_{\text{red}}$ by Theorem 9. The final question of this section is now: do both spaces even coincide or not?[14]

**4.7.1. Invariant Homomorphisms.** Before we will solve this problem in Subsection 4.8, we should first go back to the level of classical connections and re-consider the action of $S$ on parallel transports. Using the fact that parallel transports (in the sense of mappings between fibres) commute with the right action of the structure group, it is a straightforward calculation that the action of $S$ on $A \in \mathcal{A}$ gives

$$h_\gamma(\varphi_s^* A) \;=\; r^{-1} h_{s\gamma}(A) r \qquad \text{for all } s = (x,r) \in S \text{ and paths } \gamma \in \mathcal{P}.$$

As $\mathcal{A}$ is dense in $\overline{\mathcal{A}}$, the action of $S$ lifts to an action on $\overline{\mathcal{A}} = \text{Hom}(\mathcal{P}, G)$ with

$$[s \circ \overline{A}](\gamma) \;=\; r^{-1} \overline{A}(s\gamma) \, r \qquad \text{for all } s = (x,r) \in S \text{ and paths } \gamma \in \mathcal{P}.$$

Thus, we have determined the set $\overline{\mathcal{A}}_{\text{red}}$ of $S$-invariant generalized connections:

**Proposition 19.** *A generalized connection $\overline{A} \in \overline{\mathcal{A}}$ is $S$-invariant iff*

$$\overline{A}(s\gamma) \;=\; r\,\overline{A}(\gamma) r^{-1} \tag{7}$$

*for all $s = (x,r) \in S$ and all paths[15] $\gamma \in \mathcal{P}$.*

**4.7.2. Straight Lines.** Let us study how the proposition above restricts the possible parallel transports along straight lines. First of all, observe that shifting any path by some fixed $\mathbb{R}^3$ element does not alter the parallel transport at all. Thus, we may restrict ourselves to the lines $\gamma_v(t) := tv$ through the origin (w.l.o.g., $v \neq 0$). Fix now some $\overline{A} \in \overline{\mathcal{A}}_{\text{red}}$ and some $v \in \mathbb{R}^3$ and denote the parallel transport of $\overline{A}$ along $\gamma_v$ from $0$ to $t$ by $\boldsymbol{g}(t)$. Using the just obtained shift invariance and the homomorphy property of generalized connections, we see that $\boldsymbol{g} : \mathbb{R} \longrightarrow SU(2)$ is a one-parameter subgroup of $SU(2)$. Observe, however, that $\boldsymbol{g}$ need *not* be continuous. As $\mathbb{R}$ is commutative, the image of $\boldsymbol{g}$ is abelian, hence contained in a maximal torus. This means, it factorizes into $\boldsymbol{g} = T \circ \mathbf{b}$ with $\mathbf{b} : \mathbb{R} \longrightarrow S^1$ being a (possibly non-continuous) homomorphism, and $T : \mathrm{S}^1 \longrightarrow SU(2)$ (a Lie embedding of) a maximal torus in $SU(2)$. Using Pontryagin duality, we might identify $\mathbf{b}$ with an element of $\mathbb{R}_{\mathrm{Bohr}}$.

---

[14]The results of Subsections 4.7 and 4.8 can, in principle, be found in the very voluminous and interesting thesis [38] by Hanusch. Here, however, we have drastically simplified and streamlined notations, statement presentations and proofs.

[15]In order not to overload the notation, we refrain from writing "paths modulo reparametrization", here and in the following. Indeed, it should be clear that and how the action of $S$ transfers to these equivalence classes.

Can we further restrict the possible tori? Yes, we can. Observe that $\gamma_v$ is stabilized by all $r \in SU(2)$ with $rv = v$. These elements correspond to rotations $\exp(tv^i \tau_i)$ around the axis defined by $v \neq 0$. They span a maximal torus $T_v$; seen as an embedding, it is given by

$$T_v : \mathrm{S}^1 \longrightarrow SU(2) \qquad \text{with} \quad T'_v = -\mathrm{i} v^i \tau_i$$

By (7), however, this torus has to commute with all $\boldsymbol{g}(t)$. Thus, if we assume for a moment that $\boldsymbol{g}$ is not central this implies $T = T_v$. But, if $\boldsymbol{g}$ is central, we can and will choose $T = T_v$ as well. Altogether, we have $\boldsymbol{g} = T_v \circ \mathbf{b}$, and the only remaining freedom is the choice of $\mathbf{b}$.

**4.7.3. Symmetries of Paths.** It is now very desirable to learn more about the restrictions that (7) imposes on arbitrary paths. Basically, there are two constraints: first the requirement to have a homomorphism, and second to have an invariant one. For straight lines, we can see immediately their consequences: the first one implies that $\mathbf{b}$ has to be a homomorphism, the second one implies that the torus $T$ must be $T_v$. Similar statements can also be found for several other curves; see, e.g., the thesis of Hanusch [38] for a comprehensive treatment.

There, Hanusch starts his investigations by exploiting the invariance relations first. The crucial question is: which paths are in some sense independent from each other? In other words, when can a path be mapped by some element of the symmetry group $S$ to itself, or at least when do they overlap? For this, obviously, one has to investigate the symmetry behaviour of analytic paths. Basically, there are (at least) two sort-of independent types [38].

1. There are paths $\gamma$ that exhibit a continuous symmetry.          (Lie paths)
   They are given by (subpaths of) $\gamma(t) = \mathrm{e}^{tX} x$ where $X$ lies in the Lie algebra of $S$ and $x$ is in $M$. These are just the integral curves of the fundamental vector fields induced by the symmetry group action on $M$.
2. There are paths $\gamma$ that exhibit a discrete symmetry.          (brick paths)
   Here, we have (modulo the stabilizer of $\gamma$) at most finitely many elements $s \in S$ for which $s\gamma$ and $\gamma$ share a common subpath. In particular, such a $\gamma$ has a subpath that has no accumulation point with any of its $S$-translates.

Recently, the author of the present review has proven that these paths indeed comprise all possible types in the homogeneous isotropic case; even more, this statement remains true for all symmetry group actions that are pointwise proper and analytic. [32] Moreover, it is known [38] that the types above are stable w.r.t. decomposition and inversion of paths, and, of course, there is no symmetry mapping a path from one class into another. Therefore, the parallel transports can be assigned to paths in different classes completely independently (although there remain constraints within the classes), which induces a product structure on $\overline{\mathcal{A}}_{\mathrm{red}}$.

For the Lie curves, one can go even further. Indeed, also there one can introduce a notion of independence which tells us basically that the respective curves do not share full segments. Recall that any Lie curve is induced by some element in the Lie algebra $\mathfrak{s}$ of $S$. Therefore, via independence, the algebraic relations in (7) induce an equivalence relation on $\mathfrak{s}$. As non-equivalent paths can be assigned parallel transports independently, this again refines the product structure of $\overline{\mathcal{A}}_{\mathrm{red}}$. Within the factors, the problem is reduced to the constraints on parallel transports of the respective equivalence class. Finally, Hanusch obtained

$$\overline{\mathcal{A}}_{\mathrm{red}} \;=\; \mathbb{R}_{\mathrm{Bohr}} \times \left[\mathbb{R}_{\mathrm{Bohr}} \widetilde{\times} S^1\right]^{\mathbb{R}\times\mathbb{R}_+} \times \overline{\mathcal{A}}_{\mathrm{red}}^{\mathrm{brick}}, \tag{8}$$

where $\sim$ indicates some sort of projectivation. Observe that the first factor $\mathbb{R}_{\mathrm{Bohr}}$ is exactly the freedom that has remained for choosing parallel transports along straight lines (see above). The other paths require a more elaborate discussion.

### 4.8. Reduction vs Quantization

Finally, let us show that $\overline{\mathcal{A}}_{\mathrm{red}}$ is indeed larger than $\overline{\mathcal{A}_{\mathrm{red}}}$. The basic idea is to show that there are just two connections in $\overline{\mathcal{A}_{\mathrm{red}}}$ that have trivial parallel transports along all straight lines, but infinitely many in $\overline{\mathcal{A}}_{\mathrm{red}}$.

Let us begin with the parallel transports for generalized invariant connections. For this, let first $\mathbf{c} \in \mathbb{R}$ and denote the straight line connecting $0$ and $tv$ by $\gamma_t$. Recall that $\mathbf{c}$ corresponds to the smooth invariant connection $\sigma(\mathbf{c}) \equiv \theta + \mathbf{c}\eta \in \mathcal{A}_{\mathrm{red}}$, having local one-form $A = \mathbf{c}\tau_i \otimes \mathrm{d}x^i$, hence $A(\dot\gamma) = \mathbf{c}\dot\gamma^i\tau_i$. Thus the corresponding parallel transport along $\gamma_t$ is, by (3),

$$h_{\gamma_t}(\sigma(\mathbf{c})) \;=\; \mathrm{e}^{-\mathbf{c}tv^i\tau_i} \;\equiv\; T_v(\mathrm{e}^{-\mathrm{i}\mathbf{c}t}) \;\equiv\; T_v(\chi_{-t}(\mathbf{c})). \tag{9}$$

For general $\overline{\mathbf{c}} \in \overline{\mathbb{R}}$, the parallel transport of $\overline{\sigma}(\overline{\mathbf{c}}) \in \overline{\mathcal{A}}_{\mathrm{red}}$ along $\gamma_t$ is given by

$$\begin{aligned}
[\overline{\sigma}(\overline{\mathbf{c}})](\gamma_t) &= [\overline{\sigma}(\overline{\mathbf{c}})](h_{\gamma_t}) && (\overline{A}(\gamma) = \overline{A}(h_\gamma) \text{ as } \overline{\mathcal{A}} = \mathrm{Hom}(\mathcal{P}, G)) \\
&= \overline{\mathbf{c}}(h_{\gamma_t} \circ \sigma) && (\overline{\sigma}(\overline{\mathbf{c}}) = \overline{\mathbf{c}} \circ \sigma^* \text{ by Lemma 5}) \\
&= \overline{\mathbf{c}}(T_v \circ \chi_{-t}) && (\text{see (9)}) \\
&= [T_v \circ \overline{\mathbf{c}}](\chi_{-t}) && (\text{algebras extended by } \mathbb{C}^{n\times n}).
\end{aligned}$$

This shows immediately that $0 \in \mathbb{R}$ and $0_{\mathrm{Bohr}} \in \mathbb{R}_{\mathrm{Bohr}}$ are the only elements in $\overline{\mathbb{R}}$ for which the parallel transports along all straight lines are $\mathbf{1}$. (See (5) and (6).)

Let us now turn to the construction of a bunch of invariant generalized connections having trivial parallel transports along all straight lines. For every homomorphism $\mathbf{b}: \mathbb{R} \longrightarrow S^1$ define a generalized connection $\overline{A}_{\mathbf{b}}$ by

$$\overline{A}_{\mathbf{b}}(\gamma) \;:=\; \begin{cases} \mathbf{1} & \text{if } \gamma \text{ does not pass a circular arc} \\ T_n(\mathbf{b}(t)) & \text{if } \gamma \text{ has length } t \text{ and rotates in direction } n \end{cases}$$

Here, "in direction" $n$ means that the rotation axis is parallel to $n$. Moreover, the "length" $t$ is set negative if $\gamma$ rotates according the left-hand rule.[16] As $T_n$ and

---

[16] This way, rotation in direction $n$ with length $t$ equals rotation in direction $-n$ with length $-t$. The homomorphy property of $\mathbf{b}$ together with $T_{-n} = T_n^{-1}$ shows that $\overline{A}_{\mathbf{b}}(\gamma)$ is well defined.

**b** are homomorphisms, $\overline{\mathcal{A}}_{\mathbf{b}}$ is in $\overline{\mathcal{A}}$. In particular, it is trivial on all straight lines. However, it is also $S$-invariant: First, if $\gamma$ is not circular, then neither does any $s\gamma$; hence both the parallel transports on $\gamma$ and on $s\gamma$ are trivial, whence the invariance condition (7) is trivially fulfilled. Second, if $\gamma$ rotates in direction $n$ with length $t$, then $s\gamma$ rotates in direction $rn$ with length $t$, as $s = (x, r) \in S$ is an isometry. Now, (7) follows as $T_{sn} = \mathrm{Ad}_r \circ T_n$. This shows that $\overline{\mathcal{A}}_{\mathbf{b}}$ is in $\overline{\mathcal{A}}_{\mathrm{red}}$ for all **b**. And now it is clear that there are more than two connections in $\overline{\mathcal{A}}_{\mathrm{red}}$ that are trivial along straight lines. This proves

**Theorem 20.** *We have $\overline{\mathcal{A}_{\mathrm{red}}} \subset \overline{\mathcal{A}}_{\mathrm{red}}$.*

Thus we have shown for the homogeneous isotropic case:

> *Quantization and symmetry reduction do not commute.*

## 5. Conclusions

In our article we have reviewed the mathematical theory underlying the configuration space constructions of loop quantum gravity and loop quantum cosmology. Beginning with abstractly discussing the $C^*$-algebraic mathematical foundations of loop quantization in general, we have finally seen that loop quantization and symmetry reduction do not commute for homogeneous isotropic cosmology.

It is now natural to ask about other symmetries. Indeed, again Hanusch [38] has already obtained several, at least partial results for spherically symmetric, for homogeneous and for semihomogeneous cosmologies. He was able to calculate all smooth invariant connections [37]; for the spherically symmetric case, this required methods beyond the classical results by Wang [55] or Harnad, Shnider, Vinet [41]. In the other two cases, he was even able to prove again that quantization and reduction do not commute. Here, the proofs are technically more involved than for the homogeneous isotropic case. Besides, in all these cases, at least the invariant homomorphisms along Lie curves (cf. (8)) have been derived by Hanusch. This could make the definition of measures feasible not only on $\overline{\mathcal{A}_{\mathrm{red}}}$ as we showed here, but also on $\overline{\mathcal{A}}_{\mathrm{red}}$.

In general, configuration spaces is only a small part of the problem. The next one would be to study the full phase space. For loop quantum gravity, this lead to the holonomy-flux $*$-algebra and the Weyl $C^*$-algebra whose representation theory turned out to be fundamental. For homogeneous isotropic loop quantum cosmology, the result by Ashtekar and Campiglia [5] has similar relevance. However, their proof relied on the Bojowald version of the configuration space. It is now natural to ask for an extension to the framework presented here. Indeed, one can expect to get similar uniqueness as the $\mathbb{R}$-invariance already singles out the same measure as in [5], namely the Haar measure on $\mathbb{R}_{\mathrm{Bohr}}$. Having done this, one should enter the final stage of the Bojowald-Kastrup programme: the construction of symmetric quantum states.

But, even if we would be able to perform there well, there is still one conceptual issue left open. Indeed, any measure, any phase space, any uniqueness result above, rests upon one of the configuration spaces of loop quantum cosmology. But, typically, quantization and symmetry reduction do not commute. It is a completely open field to discuss all the questions above using the invariant configuration space of symmetric loop quantum gravity.

## 6. Acknowledgements

The author is very grateful to the organizers of the Regensburg conference on quantum mathematical physics for the kind invitation. The author is also very grateful to Maximilian Hanusch for numerous discussions and helpful comments on a draft version of this article.

Christian Fleischhack  
Institut für Mathematik  
Universität Paderborn  
33095 Paderborn  
Germany  
e-mail: `fleischh@math.upb.de`                    May 1, 2015